\documentclass[aps,twocolumn,showpacs]{revtex4}

\usepackage{graphicx}
\usepackage{epsfig}
\usepackage{amssymb}
\usepackage{amsmath}
\usepackage{psfrag}

\newcommand{\rg}{{\bf r}}
\newcommand{\Eg}{{\bf E}}
\newcommand{\Hg}{{\bf H}}
\newcommand{\Sg}{{\bf S}}
\newcommand{\pg}{{\bf p}}
\newcommand{\ug}{{\bf u}}
\newcommand{\jg}{{\bf j}}

\newcommand{\vg}{{\bf v}}
\newcommand{\qg}{{\bf q}}

\newcommand{\Lg}{{\bf L}}
\newcommand{\dd}{{\mathrm{d}}}
\newcommand{\ddd}{{\mathrm{d}}^2}

\begin{document}

\title{Time domain radiation and absorption by subwavelength sources}
\author{E. Bossy and R. Carminati}
\email{remi.carminati@espci.fr}
\affiliation{Institut Langevin, ESPCI ParisTech, CNRS, 10 rue Vauquelin,
75231 Paris Cedex 05, France}


\begin{abstract}
Radiation by elementary sources is a basic problem in wave physics.
We show that the time-domain energy flux radiated from electromagnetic and acoustic subwalength sources exhibits remarkable features.
In particular, a subtle trade-off between source emission and absorption underlies the mechanism of radiation. This behavior
should be observed for any kind of classical waves, thus having broad potential implications.
We discuss the implication for subwavelength focusing by time reversal with active sources.
\end{abstract}

\maketitle

%INTRODUCTION

Any textbook on wave physics or field theory contains a chapter on radiation by elementary
sources in homogeneous media~\cite{LandauCTF}. Frequency-domain analyses of the radiated fields and the
associated energy fluxes are the most widespread. In these approaches, the far-field energy flux is usually defined
as the contribution that survives time averaging, corresponding to power that continuously leaks away from the source.
Conversely, radiated near fields generate oscillating terms in the energy flux, that are discarded in the time-averaging process.
Time-domain expressions of radiated fields are also common in the context of electromagnetic radiation~\cite{StrattonBook, JacksonBook,SmithBook,Griffiths11},
including the optical regime~\cite{BornAndWolfBook}, and in acoustics~\cite{MorseBook,PierceBook,LighthillBook}.
Nevertheless, time-domain expression of the energy flux have been given much less consideration. In the case of
electromagnetic waves, the time-domain energy flux radiated from an electric dipole at rest may be found in some textbooks
(see Ref.~\cite{SmithBook} for instance). Its expression is also at the core of interesting studies of the time decay of classical
oscillating dipoles~\cite{Mandel72,Schantz95}, but that do not describe the full contribution of the
near-field terms that is discussed in the present study.
In most textbooks, the discussion is limited to harmonic oscillations and time averages, since the focus is usually on far-field
radiation~\cite{JacksonBook,BornAndWolfBook}.
In acoustics, although time-domain expressions of the fields radiated by monopole or dipole sources are widespread~\cite{MorseBook,PierceBook,LighthillBook},
we are not aware of any discussion of the time-domain energy flux, and in particular of its near-field and far-field components.

In this Letter, we revisit the basic problem of radiation by elementary subwavelength sources, from the point of view of emission
and absorption of energy in the time domain. Considering time-domain expressions of the energy flux for the acoustic monopole and
the electromagnetic dipole, and analyzing carefully the energy balance, we show that there is a subtle trade-off between emission of
energy and subsequent reabsorption by the source, the difference between emission and reabsorption giving the amount of energy
that is irreversibly radiated to the far field. This result reveals some important features of the dynamic interchange
of energy between a subwavelength source and a wavefield, that have not been discussed so far, to the best
of our knowledge. It also suggests a novel point of view on near-field radiation. Since the conclusions hold for
both acoustic and electromagnetic waves (with striking similarities), they underline a behavior that should be found with any kind
of classical waves, thus having broad implications. We illustrate an implication in the context of subwavelength focusing using time
reversal with active sources~\cite{RCPRA00,RosnyPRL02}.

%BASIC FIELD AND ENERGY FLUX EQUATIONS

The propagation of electromagnetic waves generated by a spatially localized source in an otherwise
homogeneous medium is described by the following equation~\cite{StrattonBook,JacksonBook}
\begin{equation}
\frac{1}{c^2} \frac{\partial^2 \Eg}{\partial t^2}(\rg,t) + \nabla \times \nabla \times \Eg(\rg,t) =
\Sg_{em}(\rg,t)
\label{eq:wave_EM}
\end{equation}
where $\Eg(\rg,t)$ is the electric field at point $\rg$ and time $t$,
 and $c$ is the speed of light in the medium. The source term $\Sg_{em}(\rg,t)$ is often
written in the form $\Sg_{em}(\rg,t) = -\mu_0 \, (\partial/\partial t)\jg(\rg,t)$, where $\jg(\rg,t)$ is the electric current density
and $\mu_0$ the vacuum magnetic permeability.
The electromagnetic energy current is given by  the Poynting vector ${\bf \Pi}(\rg,t) = \Eg(\rg,t) \times \Hg(\rg,t)$,
where $\Eg(\rg,t)$ is the retarded solution of Eq.~(\ref{eq:wave_EM}) and $\Hg(\rg,t)$ the associated magnetic field.
The energy flux $\phi_{em}(R,t)$ across a sphere with radius $R$ centered at the origin is
$\phi_{em}(R,t) = \int_{\mathrm{sphere}} {\bf \Pi}(\rg,t)\cdot \ug \, \ddd r$, where $\ug = \rg/|\rg|$.

For acoustic waves in the linear regime, the acoustic pressure field $p(\rg,t)$ generated by a spatially localized source
in a homogeneous medium obeys~\cite{MorseBook,PierceBook}:
\begin{equation}
\frac{1}{c_s^2} \frac{\partial^2 p}{\partial t^2}(\rg,t) - \nabla^2 p(\rg,t) =
S_{ac}(\rg,t)
\label{eq:wave_AC}
\end{equation}
where $c_s$ is the acoustic velocity in the medium and $S_{ac}(\rg,t)$ the source term.
The acoustic energy current is $\qg(\rg,t) = p(\rg,t) \vg(\rg,t)$,
$p(\rg,t)$ being the retarded acoustic pressure field solution of Eq.~(\ref{eq:wave_AC}) and $\vg(\rg,t)$ the associated
acoustic velocity field. The energy flux follows from $\phi_{ac}(R,t) = \int_{\mathrm{sphere}} \qg(\rg,t) \cdot \ug \, \ddd r$.

%SOURCE MODEL

In this Letter we study the radiation produced by sources of size much smaller than the
characteristic length scale of the wavefield, that will be denoted by ``subwavelength sources''.
In the case of electromagnetic waves, we use a point electric dipole model, with dipole moment
$\pg(t) = f(t) \pg_0$, $f(t)$ being the dimensionless time-domain amplitude and $\pg_0$ a time-independent
vector accounting for the source polarization. This model describes, e.g., a dipole moment $\pg(t) = q_e \Lg(t)$
corresponding to an oscillating charge $q_e$ with oscillation amplitude $\Lg(t)$ much smaller than all other relevant
characteristic lengths~\cite{JacksonBook}.
For a dipole centered at $\rg=0$, the electromagnetic source term
reads:
\begin{equation}
\Sg_{em}(\rg,t)=-\mu_0 \, \frac{\dd^2 \pg(t)}{\dd t^2} \, \delta(\rg)
\label{eq:source_EM}
\end{equation}
where $\delta(\rg)$ is the three-dimensional Dirac delta function.
In the case of acoustic waves, we use a point mass source model describing a radially oscillating sphere
with radius $a(t)=a_0 + \xi(t)$, in the limit of of vanishingly small radius~\cite{PierceBook}.
For a source centered at $\rg=0$, the acoustic source term reads:
\begin{equation}
S_{ac}(\rg,t) = \rho_0 \, s_0 \, \frac{\dd^2 \xi(t)}{\dd t^2} \, \delta(\rg)
\label{eq:source_AC}
\end{equation}
where $\rho_0$ is the mass density of the unperturbed homogeneous medium and $s_0=4\pi a_0^2$.
For the sake of formal similarity with the electromagnetic case, we will write $\xi(t) = f(t) \xi_0$ with
$\xi_0$ a time-independent length driving the acoustic source strength.

%EXPRESSION OF ENERGY FLUX

The time-domain solutions of Eqs.~(\ref{eq:wave_EM}) and (\ref{eq:wave_AC}) with the source terms
given by Eqs.~(\ref{eq:source_EM}) and (\ref{eq:source_AC}) can be found in textbooks on electromagnetic
and acoustic waves propagation~\cite{StrattonBook,JacksonBook,SmithBook,MorseBook,PierceBook}.
From the field expressions, the energy flux across a sphere with radius $R$ can be deduced after tedious
 but straightforward algebra. In the case of electromagnetic waves, one obtains:
 \begin{eqnarray}
\phi_{em}(R,t) &=& \frac{\mu_0 \, \pg_0^2}{6\pi \, c} \left \{ \frac{1}{2} \left ( \frac{c}{R} \right )^3 \left [ \frac{\dd f^2}{\dd t} \right ]  +
\frac{1}{2} \left ( \frac{c}{R} \right )^2 \left [ \frac{\dd^2 f^2}{\dd t^2}\right ] \right. \nonumber \\
&+&\left.  \left ( \frac{c}{R} \right ) \left [ \frac{\dd}{\dd t} \left ( \frac{\dd f}{\dd t} \right )^2 \right ]
+ \left [ \frac{\dd^2 f}{\dd t^2} \right ]^2 \right \} .
\label{eq:Phi_EM}
\end{eqnarray}
For acoustic waves, the explicit calculation of the energy flux leads to:
\begin{eqnarray}
\phi_{ac}(R,t) = \frac{\rho_0 \, s_0^2 \, \xi_0^2}{4\pi \, c_s} \left \{ \frac{1}{2} \left ( \frac{c_s}{R} \right ) \left [ \frac{\dd}{\dd t}
\left ( \frac{\dd f}{\dd t} \right )^2 \right ] + \left [ \frac{\dd^2 f}{\dd t^2} \right ]^2 \right \} .
\label{eq:Phi_AC}
\end{eqnarray}
In Eqs.~(\ref{eq:Phi_EM}) and (\ref{eq:Phi_AC}) all terms within square brackets $[ ...]$ denote retarded values,
and have to be evaluated at time $t-R/c$ (electromagnetic waves) or $t-R/c_s$ (acoustic waves).
Although their derivation is a rather simple exercise, we will see that these expressions bring to light fundamental 
aspects of the mechanism of radiation by subwavelength sources that have not been discussed so far.

% QUALITATIVE DISCUSSION (FLUX)

>From a qualitative point of view, the structure of Eqs.~(\ref{eq:Phi_EM}) and (\ref{eq:Phi_AC}) deserves
several comments. The far-field limit, obtained for $R \to \infty$, leads in both cases to an energy
flux proportional to the square of the second derivative of the source amplitude, in agreement with a well-established result
in classical wave theory~\cite{LandauCTF}. For a monochromatic source oscillating at a frequency $\omega$, with
$f(t) = \sin(\omega t)$, this far-field term is the only
one that survives a time-averaging of Eqs.~(\ref{eq:Phi_EM})  and (\ref{eq:Phi_AC}).
The far-field behavior is extensively discussed in textbooks, both for monochromatic and pulse sources.
Nevertheless the time-domain electromagnetic and acoustic energy fluxes contain additional near-field terms
whose amplitude depend on the distance $R$ to the source.
The first near-field term scales as $R^{-1}$ and is identical in Eqs.~(\ref{eq:Phi_EM}) and (\ref{eq:Phi_AC}),
except for a factor of two, while additional terms scaling as $R^{-2}$ and $R^{-3}$ appear only in the
expression for the electromagnetic case. These near-field contributions exhibit remarkable properties
that induce specific behaviors of the time-domain energy flux.
A first result is that the time-dependent amplitudes of the near-field terms in Eqs.~(\ref{eq:Phi_EM})
and (\ref{eq:Phi_AC}) read as first-order derivatives of functions that are positive (squares) and
that recover their initial values after a finite time interval (the pulse duration, or the period for monochromatic excitation).
As a result, these amplitudes necessarily change sign during their time evolution, meaning that the near-field terms lead
alternatively to outgoing or incoming contributions to the energy flux. Conversely, the far-field term only contributes
to an outgoing energy flux.
While this seems to be a commonly accepted result in the harmonic regime (for electromagnetic waves, it is known that
the Poynting vector in the near field changes sign during one cycle of oscillation), the above result precisely demonstrates
that the change of sign in the near-field energy flux also exists for a pulsed source with finite duration (i.e., with an amplitude
starting from zero and vanishing after a finite time).

% QUANTITATIVE EXAMPLES (FLUX AND ENERGY)

\begin{figure}[h]
\begin{center}
\includegraphics[width=8.6cm]{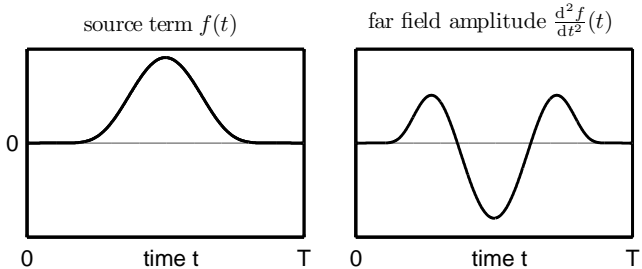}
\caption{\label{fig:amplitudes} Time evolution of the source amplitude $f(t)$ (left)
and of its second derivative $\dd^2 f(t)/\dd t^2$ (right). The latter represents the time-dependence
of the far-field amplitude for both electromagnetic and acoustic waves.}
\end{center}
\end{figure}

In order to study the behavior of the time-domain energy flux on a quantitative basis, we need to specify the source amplitude function $f(t)$.
In the present work, we consider pulses with two requirements. First, $f(t)$ has to be of strictly finite duration (denoted as $T$ in the following), in order to define exactly a pulse onset ($t=0$) and a pulse end ($t=T$). Second,  $f(t)$ and its time derivatives have to vanish continuously to zero at $t=0$ and $t=T$, in order to avoid temporal singularities in the energy flux, as made clear from Eqs.~(\ref{eq:Phi_EM}) and (\ref{eq:Phi_AC}). A broadband pulse matching these two requirement is for instance $f(t) =  \exp [ 2T^2/(t(t-T)) ] $ for $t \in ]0,T[$
and $f(t)=0$ otherwise. The temporal shape of the source amplitude $f(t)$ and the shape of the associated
far field amplitude are shown in Fig.~\ref{fig:amplitudes}. For such a broadband pulse, the period is on the order of the duration. More precisely, for the function $f(t)$ given above, the period is close to half the duration, and the corresponding wavelength is $\lambda = cT/2$ (see Fig.~\ref{fig:amplitudes}).  In this expression and in the following, for sake of brevity and since the velocities play the same role in the electromagnetic and acoustic cases, both $c$ and $c_s$ are referred to as $c$.

\begin{figure*}
\begin{center}
\includegraphics[width=18cm]{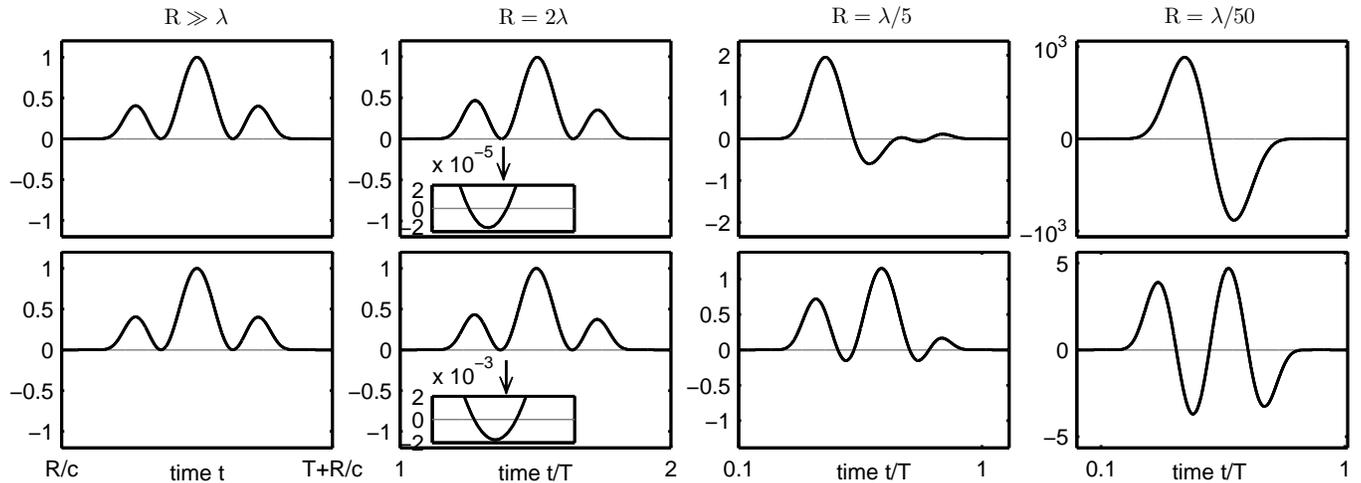}
\caption{\label{fig:flux} Time evolution of the electromagnetic and acoustic energy flux
$\phi_{em}(R,t)$ (top row) and $\phi_{ac}(R,t)$ (bottom row) for four different distance regimes.
Far-field regime $R \gg \lambda$, limit of the source free regime $R = 2\lambda$ ($2\lambda=cT$ in this particular case), near-field regimes $R <  \lambda$ and $R \ll  \lambda$. For $R=2\lambda$,
the insets show the sign inversion of the energy flux.}
\end{center}
\end{figure*}

The knowledge of $f(t)$ and its derivatives allows us to plot the time evolution of $\phi_{em}(R,t)$
and $\phi_{ac}(R,t)$ for different observation distances, covering the near-field,  the intermediate
and the far-field regimes.
We show in Fig.~\ref{fig:flux} the time evolution of the energy flux in the electromagnetic (top)
and acoustic (bottom) situations, and for four different distances.
In the far field ($R \gg \lambda$), the energy flux is always positive and describes
the radiated energy flowing irreversibly from the source. In the near field ($R \ll \lambda$), a completely
different behavior is observed. The energy flux oscillates, and takes negative values on some time intervals.
This means that part of the energy that has flowed outside the sphere of radius $R$ at a given time
flows back into the sphere at subsequent times.

\begin{figure*}
\begin{center}
\includegraphics[width=18cm]{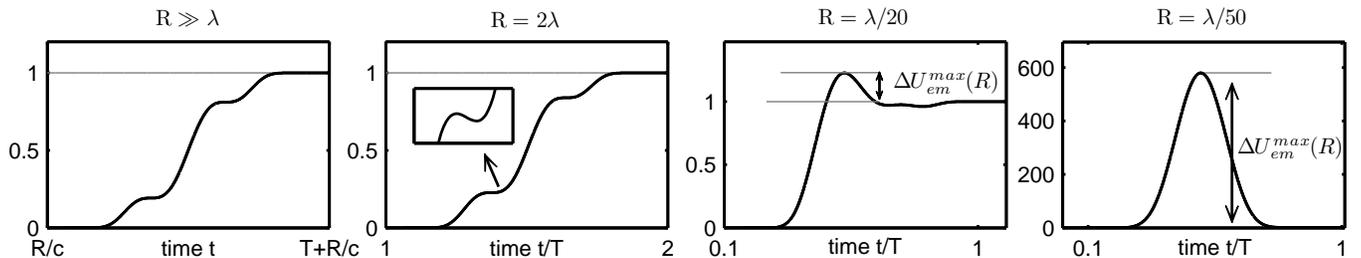}
\caption{\label{fig:energy} Time evolution of the electromagnetic energy
$U_{em}(R,t)$ stored outside the sphere of radius $R$ at time $t$,
for the same distance regimes as in Fig.~\ref{fig:flux}. The inset shows the dip due to the sign inversion
of the energy flux. }
\end{center}
\end{figure*}

At this stage, conservation of energy states that a negative energy flux corresponds to an increase of energy stored
inside the sphere with radius $R$, or to reabsorption into the source (or both). In order to quantitatively settle this point, we introduce
$U_x(R,t)$ defined as the energy stored {\it outside} the sphere with radius $R$
at time $t$ in the electromagnetic or acoustic field (the subscript ``$x$'' stands for {\it em} or {\it ac}). It reads:
\begin{equation}
U_x(R,t) = \int_0^t \phi_x(R,t^\prime) \, \dd t^\prime \ .
\label{eq:energy}
\end{equation}
The time evolution of $U_{em}(R,t)$ is shown in Fig.~\ref{fig:energy} for the same distance
regimes as in Fig.~\ref{fig:flux}. Although not shown for the sake of brevity, the same behavior
is observed for acoustic waves.
As expected from the changes in sign of the energy flux, we see that $U_{em}(R,t)$ is not a monotonic
function of time except in the far field. This non-monotonic behavior of the time evolution of the energy stored in the field can be
characterized by splitting $U_x(R,t)$ into $U_x(R,t) = U_x^\infty + \Delta U_x(R,t)$.
The first term $U_x^\infty = \int_0^\infty \phi_x(R,t) \, \dd t$ corresponds to the overall time-averaged energy
eventually radiated irreversibly through the sphere of radius $R$ to the far field, and is independent of $R$. The second term describes the time
variations of the energy stored in the field beyond the distance $R$, and either increases or
decreases $U_x(R,t)$ with respect to the asymptotic value $U_x^\infty$. This dynamic
behavior is fully described by the curves in Fig.~\ref{fig:energy}.  One clearly sees that at some time range, for $R \ll \lambda$,
the energy stored {\it outside} the sphere with radius $R$ exceeds the final energy that remains in the field after
the source has been turned off ($t>T$). This proves that part of the energy of the field has been reabsorbed by the source,
which constitute the main result of this work. This result, derived here using a pulse of finite duration and finite energy, remains valid for monochromatic
and quasi-monochromatic waves. It shows without ambiguity that a negative energy flux observed in the near field corresponds
to reabsorption by the source.

% Discussion

This conclusion puts forward new features of the near field. Although it is known that on average, near-field terms correspond to
non-radiative energy~\cite{MorseBook,NovotnyBook,reviewJJGRC}, our work shows that this non-radiative energy is dynamically exchanged between the field
and the source, at the time scale of the main oscillation. This subtle dynamic process is hidden in the first-place when computations are restricted to time-averaged values.
We also stress that a time-domain analysis reveals behaviors that cannot be seen in the frequency domain.
For example, in near-field optics or acoustics, it is often stated that some information is lost in the far field due to the loss of non-radiative components
that remain spatially localized close to the sources (in the near field zone). With a non-stationary source, one could question what happens after the source
has been turned off. Is the field finally radiated into the far field, and if so, where is the loss of information? Our work provides an unexpected answer: in the near field, some energy is constantly dynamically exchanged between the field and the source, and eventually most of it is absorbed by the source while only a small part is radiated into the far field.
The discussion has been limited in this study to a subwavelength source emitting in a homogeneous medium, so that only near fields produced
by the source itself have been considered. A more general analysis including near fields produced by scattering from subwavelength objects (secondary sources)
should also reveal interesting dynamic behaviors. In particular, understanding, in the time domain, the concept of non-radiative components that appear
in the frequency-domain angular spectrum decomposition of scattered fields~\cite{NietoBook} would be another step forward. This is left for future work.

It is also interesting to have a look
at the distance dependence  in the near field of the maximum value of the energy stored in the field
$\Delta U_{x}^{max}(R)= \mathrm{max}\{\Delta U_x(R,t)\}$. Conserving only the dominant terms
as $R \to 0$ in Eqs.~(\ref{eq:Phi_EM}) and (\ref{eq:Phi_AC}), it is easy to show that
$\Delta U_{em}^{max}(R) \sim R^{-3}$ and $\Delta U_{ac}^{max}(R) \sim R^{-1}$. Therefore for a
quasi point source model, the energy transiently  stored in the field becomes arbitrarily large
at short distance. In practice, the energy must be limited somehow by the limitations on the source model itself. Another peculiar behavior, observable only with broadband pulses of strictly finite duration, is that the energy flux exhibits a slight sign inversion even at times $t>T$, i.e., after the source has become inactive (see the insets in Fig.~\ref{fig:flux} and \ref{fig:energy}). This sign inversion does therefore not correspond to reabsorption in the source in this case, but to a small part of the energy flowing back and forth
through the sphere of radius $R$. This ``anomaly'' becomes insignificant (although non strictly zero) in the far
field since it is due to the contribution of terms in the energy flux that decay as $R^{-1}$ or faster.

To our knowledge, the near-field contributions in the time-domain energy flux have been first discussed in optics by Mandel~\cite{Mandel72}, in
the context of the decay rate of a classical electric dipole in vacuum. The discussion was constrained by the fact that for a freely decaying
atomic dipole, ``the total field energy could not exceed the maximum amount of energy of the dipole that ultimately emerges by radiation"~\cite{Mandel72}.
A major difference with the present work is that Mandel's approach considered the time variation of the envelope of the emitted wavefield,
but terms varying in time at the scale of the optical period were discarded.
In a more recent study, Schantz considered the time variations of the energy flux emitted by a decaying electric dipole keeping all time-dependent terms,
with an initial condition corresponding to an electrostatic dipole~\cite{Schantz95}. He concluded that the eventually radiated energy had
to correspond to electrostatic energy initially stored in the far field. Although it is out of the scope of this Letter to further discuss this unexpected
and interesting result, we point out that the situation studied by Schantz is very different from that considered here. Indeed,
we considered as a fundamental assumption the case of a medium initially free of energy, with a source amplitude starting exactly from zero,
and vanishing rigorously after a finite time (as opposed to an initial non-zero static field).

The results presented in this Letter were derived in the case of electromagnetic and acoustic radiation,
but they certainly underline general behaviors that should be found for any classical subwavelength sources.
Therefore, the peculiar dynamics of the energy exchange between a subwavelength source and the radiated field has potentially broad implications.
Here, we discuss one important consequence in the context of subwavelength focusing by time reversal. Experimental realizations of time-reversed wavefields have been demonstrated both in acoustics and electromagnetism, by use of close 2D or 3D cavities~\cite{DraegerPRL97,RosnyPRL02,LeroseyScience07}. When the field emitted by a point-like source is time-reversed in the source-free medium, refocusing is limited by diffraction~\cite{DraegerPRL97}. However, when both the wavefield and the source are time reversed, perfect refocusing can be obtained~\cite{RCPRA00,NietoBook}. Accordingly, experiments in acoustics have demonstrated subwavelength refocusing with an active time-reversed
source~\cite{RosnyPRL02}, the focal spot size being limited only by the finite size of the source itself.
Intuitively, the role of the time-reversed source is seen as that of a sink, i.e., of an absorber of the incoming time-reversed wave.
Our work shows that the role of the time-reversed source is more subtle, and that it necessarily involves both absorption and emission of energy.
Indeed, the time-domain evolution of the field energy in a perfect time-reversal experiment (with reversed field and source) is directly given by the curves in Fig. 3
read backwards. Therefore, the energy in the field is transiently larger than the energy carried by the time-reversed wavefield, so that in some time range, the sink actually
behaves as a source. The time-reversed source is both an absorber and an emitter. The term "sink" therefore only makes sense when one considers the
overall energy balance, obtained after time integration. Our work has two important consequences for practical experiments. First, the focusing performances
cannot be discussed without considering the energy point of view, in particular because for a sink of vanishingly small size, the transient energy that
has to be stored in the field becomes arbitrarily large. Second, perfect subwavelength refocusing (i.e., without energy scattered away from the focal spot)
cannot be achieved by use of a passive subwavelength absorber, as efficient as it may be, since the dynamic exchange of energy is a necessary condition for
a localized absorption of the full energy of the wavefield.

% CONCLUSION

In summary, from the study of time-domain expression of the energy flux radiated by pulsed electromagnetic and acoustic elementary sources, we have shown that the non-radiative energy predominant in the near-field is dynamically exchanged between the source and the field.
We have discussed implications for subwavelength focusing and imaging. Since the results hold for both electromagnetic and acoustic waves, we believe
that they underly a universal process of radiation by any kind of subwavelength sources, although demonstrated here only for the acoustic monopole and the electromagnetic dipole. In the case of
electromagnetic waves emitted by a single classical  dipole emitter, a giant transient storage of
electromagnetic energy is necessary in order to radiate a (much smaller part) in the far field.
It would be interesting to clarify the way quantum theory handles this point in the computation
of spontaneous emission by a single atom.

We acknowledge A.C. Boccara, J.J. S\'aenz and A. Sentenac for helpful discussions.
This work was supported by the Agence Nationale de la Recherche (grant JCJC07-195015).

%%%%%%%%%%%%%%

\end{document}